\documentclass[a4paper,11pt]{article}
\pdfoutput=1
\usepackage{typearea}
\areaset[0cm]
        {16cm}{23cm}         
\usepackage[latin1]{inputenc}
\usepackage[T1]{fontenc}
\usepackage{amsmath}
\usepackage{amssymb}
\usepackage{mathcomp}
\usepackage{graphicx}
\usepackage{epstopdf}
\usepackage{colortbl}
\usepackage{bbold}

\usepackage{slashed}

\definecolor{head}{gray}{0.9}
\newcolumntype{A}{%
>{\columncolor{head}}l}

\usepackage[small,bf]{caption}
\newcommand{\mpl}{M_{\rm Pl}} 
\newcommand{\phie}{\phi_{\rm e}}
\newcommand{\phis}{\phi_*}
\newcommand{\braket}[1]{\langle #1 \rangle}
\newcommand{\phimin}{\phi_{\rm min}}
\newcommand{\dphi}{\operatorname{d\phi}}
\newcommand{\del}{\delta \phi}

\begin{document}

\begin{titlepage}

\vspace*{-15mm}
\vspace*{0.7cm}

\begin{center}

{\Large {\bf BICEP2 implications for single-field slow-roll inflation revisited}}\\[8mm]

Stefan Antusch$^{\star\dagger}$\footnote{Email: \texttt{stefan.antusch@unibas.ch}} and
David Nolde$^{\star}$\footnote{Email: \texttt{david.nolde@unibas.ch}}

\end{center}

\vspace*{0.20cm}

\centerline{$^{\star}$ \it
Department of Physics, University of Basel,}
\centerline{\it
Klingelbergstr.\ 82, CH-4056 Basel, Switzerland}

\vspace*{0.4cm}

\centerline{$^{\dagger}$ \it
Max-Planck-Institut f\"ur Physik (Werner-Heisenberg-Institut),}
\centerline{\it
F\"ohringer Ring 6, D-80805 M\"unchen, Germany}

\vspace*{1.2cm}

\begin{abstract}
\noindent
It is generally believed that in single-field slow-roll inflation, a large tensor-to-scalar ratio $r > 0.1$ requires inflaton field values close to or above the Planck scale. Recently, it has been claimed that $r > 0.15$ can be achieved with much smaller inflaton field values $\Delta\phi < \mpl/10$. We show that in single-field slow-roll inflation, it is impossible to reconcile $r > 0.1$ with such small field values, independently of the form of the potential, and that the recent claim to the contrary is based on an invalid approximation. We conclude that the result of the BICEP2 measurement of $r > 0.1$, if confirmed, truly has the potential to rule out small-field models of single-field slow-roll inflation.
\end{abstract}
\end{titlepage}

\newpage

\section{Introduction}

A few weeks ago, the BICEP2 collaboration reported a discovery of B-mode polarization of the CMB \cite{BICEP}, for which the most obvious explanation would be primordial gravity waves due to vacuum fluctuations of the metric during inflation. Though it is still too early to be sure about this interpretation, the announcement has already generated great excitement due to the many dramatic implications such a detection of primordial tensor perturbations would have \cite{LythBound1,LythBound2,LythBound3,LythBound4,implication1,implication2}.

One of these implications is that in slow-roll inflation, a large tensor-to-scalar ratio $r$ requires values of the inflaton field close to or above the Planck scale. The basic idea, known as the Lyth bound \cite{LythBound1,LythBound2,LythBound3,LythBound4}, is that a large tensor-to-scalar ratio requires a steep inflaton potential at horizon crossing, and that in such a steep potential, the inflaton moves over a large field range during the 50-60 e-folds of inflation that are needed to solve the flatness and horizon problems.

The argument is often presented for a monotonous slow-roll parameter $\varepsilon$, for which this bound is particularly strong. It is therefore natural to ask whether the Lyth bound can be circumvented by a non-monotonous evolution of $\varepsilon$. In fact, it has been recently claimed that a large $r > 0.15$ could be achieved in a single-field slow-roll model with $\Delta \phi < 0.1 \mpl$ \cite{LythBoundViolation1,LythBoundViolation2}. In this letter we show that requiring single-field slow-roll inflation alone is sufficient to rule out small-field inflation for the large tensor-to-scalar ratio $r \gtrsim 0.1$ implied by the BICEP2 experiment. As an illustrative example, we also explain how the Lyth bound is enforced for the specific inflaton potential used in \cite{LythBoundViolation1}. We close with a summary and a brief discussion of the implications of our result.

\section{Lyth bound revisited}

Throughout this letter, we generally work in units with $\mpl = 1$. However, we sometimes write $\mpl$ explicitly to emphasize the mass dimension in some equations.

\subsection*{Simplest form of the Lyth bound}

We start with a brief review of the Lyth bound in its simplest form, assuming for now not only single-field slow-roll inflation, but also a roughly constant slow-roll parameter $\varepsilon$, where the slow-roll parameters are defined as \cite{slowroll1}
\begin{align}
 \varepsilon \, = \, \frac{1}{2} \left( \frac{V'(\phi)}{V(\phi)} \right)^2, \quad
 \eta \, = \, \frac{V''(\phi)}{V(\phi)}, \quad
 \xi^2 \, = \, \frac{V'(\phi)V'''(\phi)}{V^2(\phi)}, \quad
 \sigma^3 \, = \, \frac{[V'(\phi)]^2 V''''(\phi)}{V^3(\phi)}.
 \label{eq:slowrollParams}
\end{align}
The number of e-folds realized during inflation, in leading order of the slow-roll approximation, is given by
\begin{align}
 N_{\rm tot} \, = \, \int\limits_{\phie}^{\phis} \dphi \frac{V(\phi)}{V'(\phi)} \, < \, \lvert \phis - \phie \rvert \left| \frac{V}{V'} \right|_{\rm max} \, = \, \lvert \phis - \phie \rvert \frac{ 1 }{ \sqrt{2\varepsilon_{\rm min}} }, \label{eq:simpleLyth1}
\end{align}
where the index $*$ denotes the time when CMB scales cross the horizon and the index ``e'' denotes the end of inflation. Eq.~\eqref{eq:simpleLyth1} implies a bound on the field value:
\begin{align}
 \Delta \phi \, \equiv \, \lvert \phis - \phie \rvert \, > \, N_{\rm tot} \sqrt{2\varepsilon_{\rm min}}.
\end{align}
If we assume that $\varepsilon_{\rm min} \sim \varepsilon_*$, we can replace $\varepsilon$ by the tensor-to-scalar ratio $r$ using
\begin{align}
 r \, = \, 16\varepsilon_*
\end{align}
to arrive at the usual Lyth bound
\begin{align}
  \Delta \phi \, \sim \, N_{\rm tot} \sqrt{r/8}. \label{eq:classicalLythResult}
\end{align}
With $N_{\rm tot} \sim 50$ and $r \gtrsim 0.1$, this implies $\Delta \phi \gtrsim 5\mpl$.

\subsection*{Lyth bound for general single-field slow-roll inflation}

\begin{figure}[tbp]
  \centering
\includegraphics[width=0.48\textwidth]{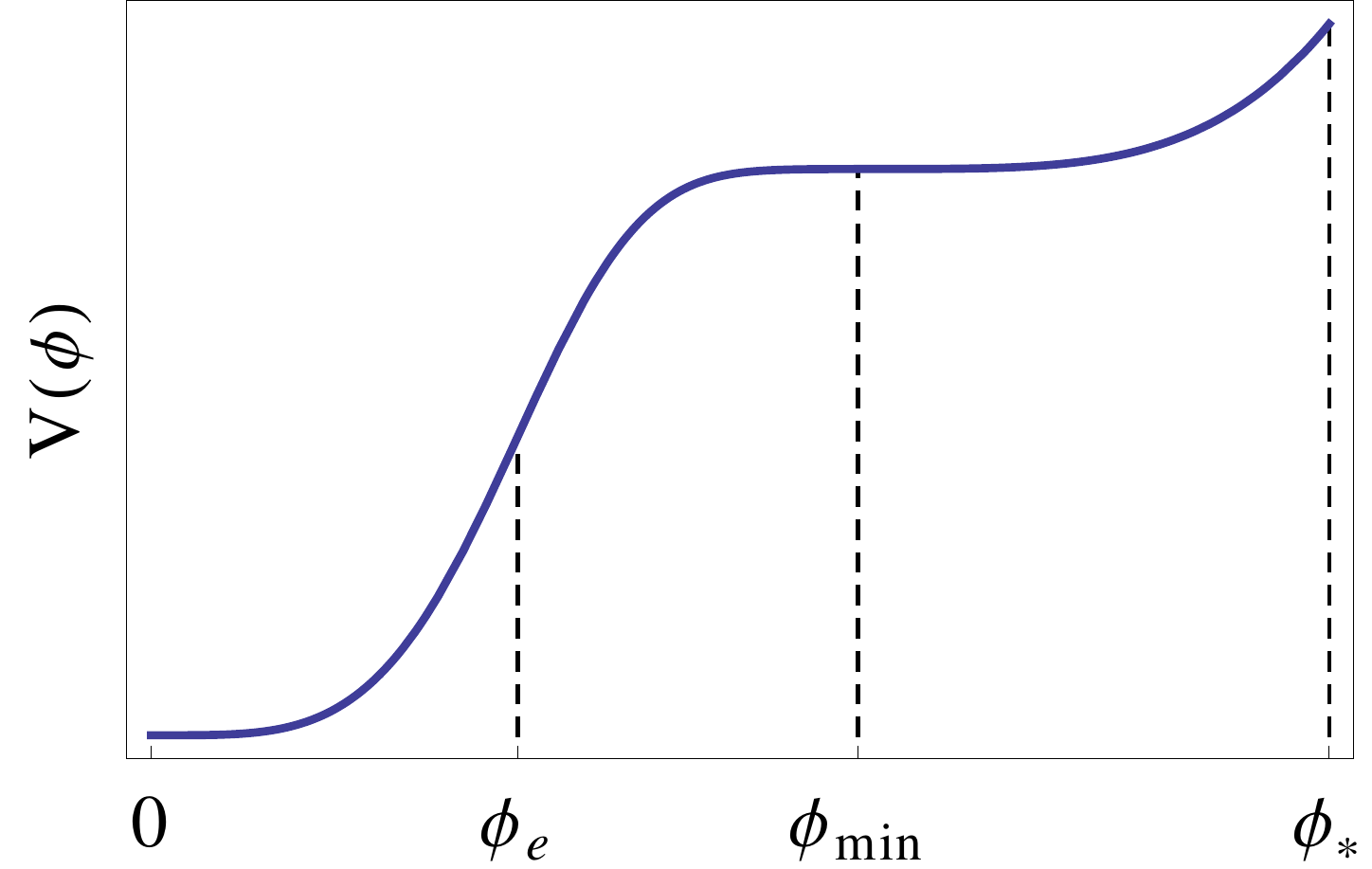}
  \caption{Schematic potential shape which would be optimal for evading the Lyth bound. The slope is very big at the horizon crossing scale $\phis$, thereby generating the required tensor-to-scalar ratio $r=16\varepsilon_*$. It then quickly flattens until a point $\phimin$ where the slow-roll parameter $\varepsilon$ has a local minimum. This flat region is necessary to generate sufficiently many e-folds with small field excursion, as $\operatorname{dN/d\phi} \simeq (2\varepsilon)^{-1/2}$. Afterwards, inflation ends at some point $\phie \leq \phimin$, either because the potential becomes to steep for slow-roll inflation or because of a waterfall transition like in hybrid inflation.}
  \label{fig:skizzePotential}
\end{figure}

Looking at the derivation leading to eq.~\eqref{eq:classicalLythResult}, one quickly realizes that the bound can be circumvented if $\varepsilon_{\text{min}} \ll \varepsilon_*$. The Lyth bound arises because we need large $\varepsilon_*$ to get a large $r$, but getting enough e-folds with small inflaton field values requires large $dN/d\phi = (2\varepsilon)^{-1/2}$ and therefore small $\varepsilon(\phi)$. One might therefore think that the Lyth bound from eq.~\eqref{eq:classicalLythResult} can be circumvented if $\varepsilon$ is only large at $\phi_*$ and very quickly goes to small values for $\phi < \phi_*$, as shown in fig.~\ref{fig:skizzePotential}.

This actually works up to a point, and the Lyth bound can be weakened. However, in slow-roll inflation, $\varepsilon$ cannot vary arbitrarily quickly. In this section, we will show that this constraint makes it impossible to have single-field slow-roll inflation with large $r \gtrsim 0.1$ and $\Delta\phi \ll \mpl$, independently of the form of the potential.

We define $\phi_*$ as the field value at horizon crossing, $\phie$ as the field value at the end of inflation, and $\phimin$ as the field value where $\varepsilon$ has its minimum in the interval between $\phi_*$ and $\phie$ (see fig.~\ref{fig:skizzePotential}). We can assume that $\phie < \phi_*$, so that $\phie \leq \phimin \leq \phi_*$.\footnote{If this is not the case for some inflaton field $\tilde{\phi}$, we can always perform a field redefinition $\tilde{\phi} = \tilde{\phi}_\text{e} - \phi$, so that our assumption is correct for the field $\phi$.}

The definitions of the slow-roll parameters in eq.~\eqref{eq:slowrollParams} imply that
\begin{align}
 \operatorname{ \frac{d}{d \phi} } \sqrt{2\varepsilon} \, = \, \operatorname{ \frac{d}{d \phi} } \frac{ V' }{ V } \, = \, \frac{ V'' }{ V } - \left( \frac{V'}{V} \right)^2 \, = \, \eta - 2\varepsilon. \label{eq:epsilonDerivative}
\end{align}
During slow-roll, one has $\varepsilon \ll 1$ and $\eta \ll 1$. Therefore, eq.~\eqref{eq:epsilonDerivative} tells us that the first slow-roll parameter $\varepsilon$ cannot change too quickly during slow-roll inflation.

Integrating eq.~\eqref{eq:epsilonDerivative} over $\phi$ from $\phimin$ to $\phi_*$, we find
\begin{align}
 \left( \frac{V'}{V} \right)_* - \left( \frac{V'}{V} \right)_{\text{min}}    \, = \, \int\limits_{\phimin}^{\phi_*}\dphi \operatorname{ \frac{d}{d \phi} } \sqrt{2\varepsilon} \, = \, \int\limits_{\phimin}^{\phi_*}\dphi \left( \eta - 2\varepsilon \right) \, = \, \left( \phi_* - \phimin \right) \braket{ \eta - 2\varepsilon }, \label{eq:epsilonDerivativeIntegrated}
\end{align}
where $\braket{ \eta - 2\varepsilon }$ is the mean of $(\eta - 2\varepsilon)$ between $\phimin$ and $\phi_*$.

We next want to show that $(V'/V)_* \gg (V'/V)_\text{min}$. For large $r = 16\varepsilon_*$, we have
\begin{align}
 \left( \frac{V'}{V} \right)_* \, = \, \sqrt{\frac{ r }{ 8 }} \, \simeq \, 0.11 \times \sqrt{\frac{r}{0.1}}. \label{eq:VstarBound}
\end{align}
A bound on $(V'/V)_\text{min}$ can be derived from
\begin{align}
 \phi_* - \phie \, = \, \int\limits_{N_{\rm e}}^{N_*}\operatorname{dN} \operatorname{\frac{d\phi}{dN}} \, = \, \int\limits_{N_{\rm e}}^{N_*}\operatorname{dN} \frac{V'(\phi)}{V(\phi)} \, > \, N_\text{total} \left( \frac{ V' }{ V } \right)_\text{min}.
\end{align}
With $N_\text{total} \sim 50$ and inflaton field values $\Delta \phi = \phi_* - \phie$, we get the bound
\begin{align}
 \left( \frac{ V' }{ V } \right)_\text{min} \, < \, \frac{\phi_* - \phie}{N_\text{total}} \, \lesssim \, 0.02 \left( \frac{\Delta \phi}{\mpl} \right). \label{eq:VminBound}
\end{align}

For large $r \gtrsim 0.1$ and $\Delta \phi \lesssim \mpl$, we can neglect $(V'/V)_\text{min}$ in eq.~\eqref{eq:epsilonDerivativeIntegrated}, as it is much smaller than $(V'/V)_*$. We then plug in eq.~\eqref{eq:VstarBound} and use $\Delta \phi \geq \phi_* - \phimin$ to find our main result
\begin{align}
 \frac{\Delta \phi}{\mpl} \, \gtrsim \, \frac{ 0.11 }{\braket{\eta - 2\varepsilon}} \sqrt{\frac{r}{0.1}}. \label{eq:MAIN}
\end{align}
Eq.~\eqref{eq:MAIN} is the main equation of this letter, as it clearly shows that generating $r \gtrsim 0.1$ from single-field slow-roll inflation requires inflaton field excursions close to or above the Planck scale (recall that both $\varepsilon$ and $\eta$ must be small during slow-roll inflation). The derivation did not rely on any assumption except on the validity of the slow-roll approximation and $\Delta \phi \lesssim \mpl$, so this form of the Lyth bound cannot be circumvented by any particular choice of slow-roll inflaton potential, even if the evolution of $\varepsilon$ is not monotonous.

We want to mention that if one requires slow-roll inflation only throughout the $\Delta N \sim 8$ e-folds of inflation for which the primordial spectrum has been probed by CMB observations,\footnote{In this case, the remaining $(N_{\text{tot}} - \Delta N)$ e-folds must be generated by another mechanism, e.g. a second phase of slow-roll inflation or thermal inflation.} the bound on $(V'/V)_{\rm min}$ in eq.~\eqref{eq:VminBound} weakens by a factor of $6$. However, for $\Delta \phi \ll \mpl$ there will still be a suppression by $\Delta \phi/\mpl$. Therefore, one still has $(V'/V)_{\rm min} \ll (V'/V)_*$ for small-field models. Only for $\Delta\phi \sim \mpl$, the prefactor in eq.~\eqref{eq:MAIN} will change, but our conclusion that $r \gtrsim 0.1$ requires $\Delta \phi \gtrsim \mpl$ remains unchanged.

\subsection*{Remark on multi-field models}

Note that the arguments leading to eq.~\eqref{eq:MAIN} are also valid for multi-field models if one replaces $V'$ and $V''$ by the derivatives of $V$ along the field trajectory and $\Delta \phi$ by the length of the trajectory in field space. However, one can in principle avoid the conclusion that individual fields must have super-Planckian excursions. There are two possibilities to realize this. The first one is to have very curved field trajectories that spiral \cite{spiralMultifield} or zig-zag within a small sphere with radius $\phi_{\rm max} \ll \mpl$, so that the length of the trajectory is much larger than the radius $\phi_{\rm max}$. As the slow-roll trajectory always moves along the gradient of the potential, this requires very non-trivial scalar potentials. The other possibility is to use a large number $\mathcal{N}$ of inflaton fields \cite{Nflation1}; then by the Pythagorean theorem, each field needs only needs to travel a shorter distance $\Delta \phi_i \sim \Delta \phi / \sqrt{ \mathcal{N} } \ll \Delta \phi$.

\subsection*{Remark on canonical normalization}

The arguments used above assume that the inflaton field $\phi$ is canonically normalized. We think that when talking about large- and small-field models, it is useful to use the canonically normalized inflaton field, as this makes the categorization independent of arbitrary field redefinitions. Otherwise, every model could be interpreted as a ``small-field'' model by e.g.\ replacing the inflaton field $\phi$ with a rescaled inflaton field $\tilde{\phi} = \phi / (\alpha \Delta \phi)$ with $\alpha \gg 1/\mpl$. For single-field models with non-canonical kinetic terms, eq.~\eqref{eq:MAIN} is applicable only after the inflaton field has been redefined to have canonical kinetic terms.

\section{Explicit bounds from inflaton potential reconstruction}

To get a better understanding of the bound from eq.~\eqref{eq:MAIN}, we show how a local reconstruction of the inflaton potential from the CMB observables forces us towards Planckian field values. Note that this reconstruction only fixes the potential for field values $\phi \sim \phi_*$, so it only provides a good approximation to the full inflaton potential for small-field models. Our approach in this section is to assume that we could achieve $r \gtrsim 0.1$ in a small-field model, write down the expansion of the inflation potential, and show that this leads us to require large field values.

Assuming a small-field model, we expand the potential around $\phi_*$, defining $\del = (\phi - \phi_*)$:
\begin{align}
 V(\del) \, &= \, V_*\left( 1 + \frac{V'_*}{V_*} \, \del + \frac{V''_*}{2V_*} \, \del^2 + \frac{V'''_*}{6V_*} \, \del^3 + \frac{V''''_*}{24V_*} \, \del^4 \right) + ... \notag\\
 &= \, V_* \left( 1 + \sqrt{2\varepsilon_*} \, \del + \frac{\eta_*}{2} \, \del^2 + \frac{\xi^2_*}{6\sqrt{2\varepsilon_*}} \, \del^3 + \frac{\sigma^3_*}{48 \varepsilon_*} \, \del^4 \right) + ...,  \label{eq:VAnsatz}
\end{align}
with the slow-roll parameters as defined in eq.~\eqref{eq:slowrollParams}.\footnote{There are two possible choices for the sign of $(V'/V) = \pm\sqrt{2\varepsilon}$. These are physically equivalent, as one can always switch between the two by a field redefinition $\delta\phi \rightarrow -\delta\phi$.}

The primordial spectrum from slow-roll inflation can be calculated from the slow-roll parameters at $\phi_*$. These relations can be solved for $V_*$ and the slow-roll parameters at $\phi_*$ depending on the primordial spectrum parameters $A_s$, $r$, $n_s$, $\alpha_s$ and $\kappa_s$. To lowest order in the slow-roll approximation, one finds
\begin{align}
 \varepsilon_* \, \simeq \, \frac{r}{16}, \quad~
 \eta_* \, \simeq \, \frac{n_s - 1}{2} + \frac{3r}{16}, \quad~
 \xi^2_* \, \simeq \, - \frac{\alpha_s}{2} + ..., \quad~
 \sigma^3_* \, \simeq \, \frac{\kappa_s}{2} + ...,
\end{align}
where the ... denote products of observables which are roughly of order $10^{-3}$. Using the bounds from the Planck satellite \cite{Planck}, one finds that none of the slow-roll parameters $\eta$, $\xi^2$, $\sigma^3$ can be larger than $O(10^{-2})$ at the pivot scale $\phi_*$.

\begin{figure}[tbp]
  \centering
$\begin{array}{cc}
\includegraphics[width=0.48\textwidth]{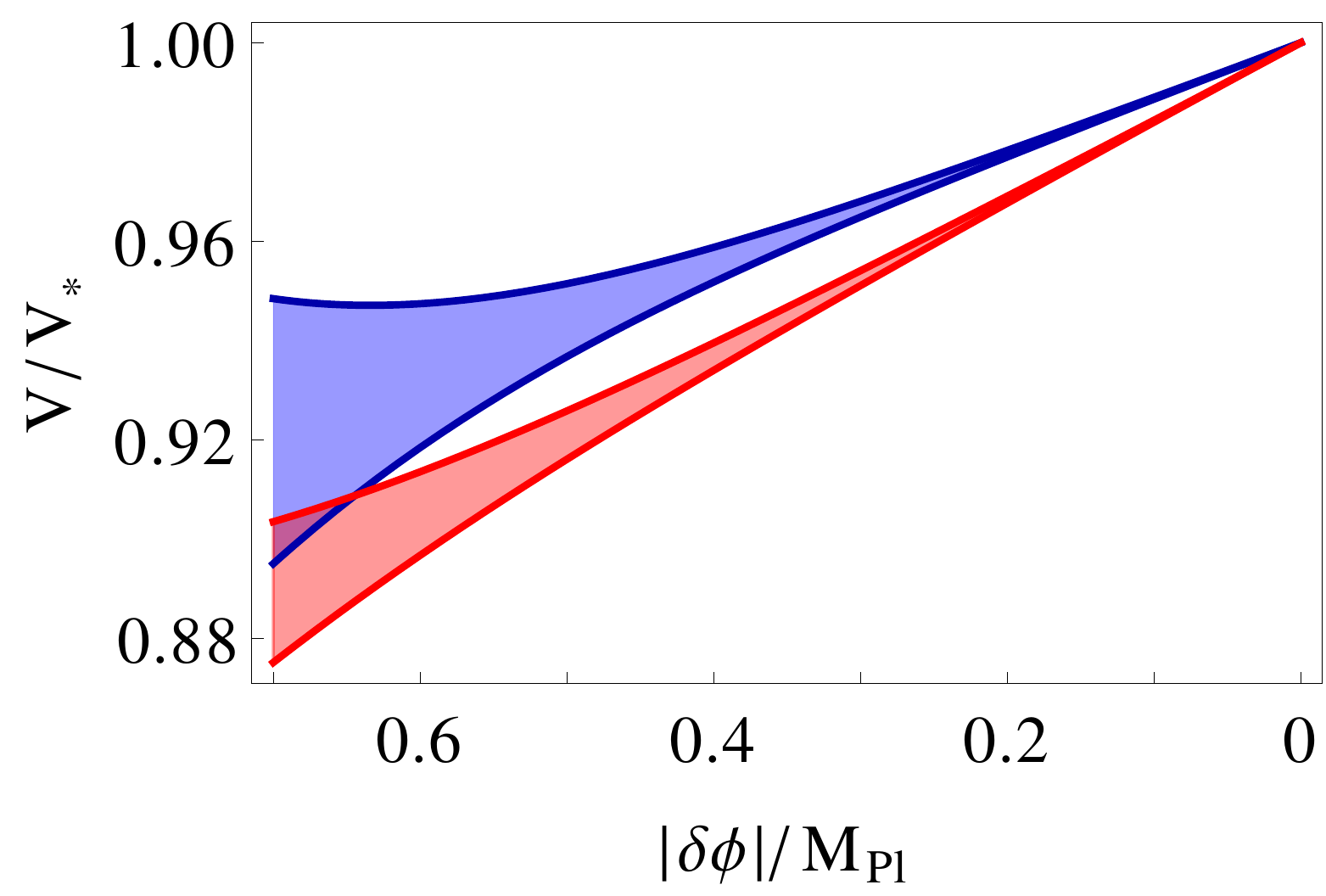} &
\includegraphics[width=0.48\textwidth]{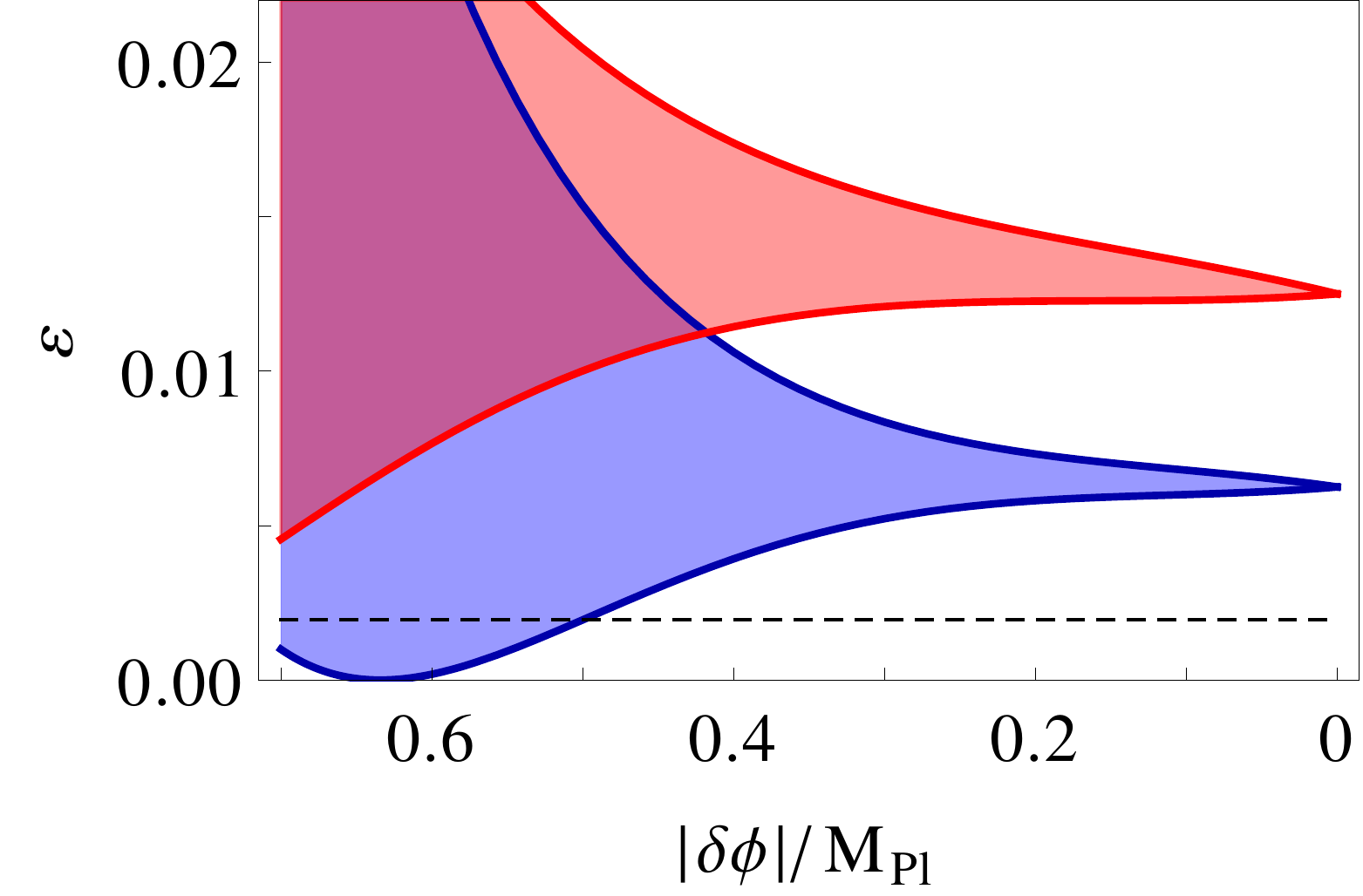}
\end{array}$
\caption{\textit{Left:} local reconstruction of the inflaton potential around $\phi_*$ for $r=0.1$ (blue band) and $r=0.2$ (red band), see eq.~\eqref{eq:VAnsatz}. The width of the bands (upper and lower bound of the potential) is due to the uncertainty in the slow-roll parameters $\eta$, $\xi^2$ and $\sigma^3$, which must not be larger than $O(10^{-2})$ in order to satisfy the Planck constraints on the observed approximate scale invariance of the scalar spectrum. For this plot, we have allowed these slow-roll parameters to vary within the range $[-0.05,0.05]$. Due to the large value of $r$, the potential is then forced to be nearly linear near $\phi_*$. \newline \textit{Right:} Slow-roll parameter $\varepsilon$ for this potential. Because the potential is nearly linear near $\phi_*$, $\varepsilon$ cannot decrease significantly until $\lvert \delta \phi \rvert > \mpl/2$. This makes it impossible to generate enough e-foldings of inflation with small field values. As a reference, the maximum $\varepsilon$ for which $\Delta N \sim 8$ e-folds could be realized within $\Delta \phi \leq \mpl/2$ is shown as a dashed horizontal line.}
  \label{fig:potentialReconstruction}
\end{figure}

In the left plot of fig.~\ref{fig:potentialReconstruction}, we show the upper and lower bound of the potential if we allow $\eta_*$, $\xi^2_*$ and $\sigma^3_*$ to vary in the interval $[-0.05,0.05]$, while fixing $\varepsilon_*$ by setting $r = 0.1$ (blue lines) or $r = 0.2$ (red lines). In the right plot of fig.~\ref{fig:potentialReconstruction}, we see the corresponding slow-roll parameter $\varepsilon$. We see that the potential is close to linear up to $\lvert \delta \phi \rvert \sim 0.4\mpl$, and we cannot recover a potential of the form shown in fig.~\ref{fig:skizzePotential}, which we would need to evade the Lyth bound.

What happened here? If we take a look at eq.~\eqref{eq:VAnsatz}, we see that the coefficient of the linear term is proportional to $\sqrt{2\varepsilon_*} = \sqrt{r/8}$, whereas the higher-order terms are suppressed by powers of $\sqrt{r}$. As the other slow-roll parameters cannot be larger than $O(10^{-2})$, this means that the potential becomes more and more dominated by the linear term when one increases $r$. However, for a linear potential with a large, slowly varying vacuum energy $V$, it is clear that $\varepsilon =\frac{1}{2}(V'/V)^2$ is nearly constant, which means that the conditions of the traditional Lyth bound in eq.~\eqref{eq:classicalLythResult} are satisfied.

\subsection*{Comparison to reconstructions of the inflaton potential in the literature}

This kind of potential reconstruction has been discussed carefully, using the constraints available at that time, in a number of insightful earlier papers \cite{smallField1,smallField2} (see \cite{potentialReconstructionRecent} for recent work including the BICEP2 data). They find that having a large tensor-to-scalar ratio $r$ with $\Delta \phi \lesssim \mpl$ forces the higher-order slow-roll parameters $\xi^2$ and $\sigma^3$ to become large -- which they see numerically as a deviation of the primordial spectrum from the usual power law --, so that higher order slow-roll corrections should be taken into account, particularly for the spectral index $n_s$. They generally do not find that $\Delta \phi \ll \mpl$ is possible: their field values lie in the range $\mpl/2 \lesssim \Delta \phi \lesssim \mpl$, which is already too large to reliably use expansions in powers of $\phi/\mpl$.

These results fit very well to our general bound in eq.~\eqref{eq:MAIN}: one can arrange for $r \simeq 0.1$ with field values slightly below the Planck scale, but in order to do so, one needs to deviate from perfect slow-roll, for which reason higher order slow-roll corrections to the observables become relevant. This is just what one would expect from our bound $\Delta \phi \braket{\eta - 2 \varepsilon} \gtrsim 0.11 \mpl$, as pushing the field range below the Planck scale will inevitably require larger slow-roll parameters.

\subsection*{Recent claim of $r > 0.15$ with $\phi < 0.1 \mpl$}

Our results contradict the recent claim \cite{LythBoundViolation1} that a large tensor-to-scalar ratio $r > 0.15$ could be achieved with small field values $0.066 \leq \frac{\Delta \phi}{\mpl} \leq 0.092$ in just the kind of potential reconstruction we have sketched above. Their approach differs from ours in that they derive $\Delta \phi$ not by integrating $dN/d\phi = V(\phi)/V'(\phi)$ over $\phi$, but by an integration in momentum space (eq.~(2.6) in \cite{LythBoundViolation1}), which is of the form:
\begin{align}
 \frac{\Delta \phi}{\mpl} \, \simeq \, \sqrt{\frac{r}{8}} \int\limits_{k_e}^{k_*} \frac{ \operatorname{dk} }{ k } \left(  \frac{k}{k_*}  \right)^{\frac{a}{2} + \frac{b}{4} \ln\left( \frac{k}{k_*} \right) + \frac{c}{12}\ln^2\left( \frac{k}{k_*} \right) }, \label{eq:phiIntegral}
\end{align}
where they integrate over $\Delta N \sim 17$ e-folds, so that $k_e/k_* \sim \exp(-17)$, and $a$, $b$ and $c$ are the spectral index, running, and running of the running of $r(k)$:
\begin{align}
 a = n_t - n_s + 1, \quad b = \alpha_t - \alpha_s, \quad c = \kappa_t - \kappa_s.
\end{align}
They use an approximate solution to this integral given in eqs.~(4.4) and (4.5) in their previous paper \cite{LythBoundViolation2}, which is
\begin{align}
 \int\limits_{k_e}^{k_*} \frac{ \operatorname{dk} }{ k } \left(  \frac{k}{k_*}  \right)^{\frac{a}{2} + \frac{b}{4} \ln\left( \frac{k}{k_*} \right) + \frac{c}{12}\ln^2\left( \frac{k}{k_*} \right) } \, \simeq \, \left(  \frac{a}{4}-\frac{b}{16}+\frac{c}{48}-\frac{1}{2}  \right) + ... \quad~~~\text{(for $a>b>c \neq 0$)}
\end{align}
However, for small $a$, $b$ and $c$, as one predicts from slow-roll inflation, this approximation is actually very bad. In fact, when comparing with the limit $a$, $b$, $c \rightarrow 0$, for which the integral is trivial, one finds that the approximation underestimates the true value of the integral by a factor of $2\Delta N$. We have checked numerically that the approximation is also wrong by a factor of $\sim 10-30$ for most values of $a > b > c$ with $a \lesssim O(10^{-2})$, which would be a realistic value for $a$ during slow-roll inflation.

As $\Delta \phi$ is approximately proportional to this integral, this underestimates $\Delta \phi$ roughly by a factor of $10-30$, which brings their actual inflaton field value up to $\Delta \phi = O(\mpl)$. As both papers \cite{LythBoundViolation1,LythBoundViolation2} neglect higher powers of $\Delta \phi/\mpl$ in their calculation (because they assume to have a small-field model), their calculation has to be revised. However, as we have shown above, it is generally impossible to realize $r \gtrsim 0.1$ in any small field-model of single-field slow-roll inflation, so one will find that for any such model $\phi \ll \mpl$ implies $r < 0.1$.

\section{Summary and conclusions}

In this letter, we have discussed why it is impossible to construct a single-field slow-roll model of small-field inflation (defined as $\Delta \phi \ll \mpl$) that generates $r \gtrsim 0.1$. Making no assumptions beyond single-field slow-roll inflation,\footnote{To be precise, the derivation also assumed that $\Delta \phi \lesssim \mpl$, because our bound is aimed at ruling out small-field models. For super-Planckian field values it is anyway clear that one can generate large $r$.} we arrived at the bound \eqref{eq:MAIN}
\begin{align*}
 \Delta \phi \, \gtrsim \, \frac{ 0.11 }{\braket{\eta - 2\varepsilon}} \sqrt{\frac{r}{0.1}},
\end{align*}
which requires that for slow-roll inflation (for which $\varepsilon \ll 1$ and $\eta \ll 1$), the canonically normalized inflaton field must take values close to or above the Planck scale. We also briefly discussed how this bound can be generalized to multi-field slow-roll models, and mentioned two ways in which multi-field models can dodge this bound.

We demonstrated that our results hold up for an expansion of the inflaton potential to 4th order, for which it was claimed that large $r > 0.15$ could be realized in small-field slow-roll inflation \cite{LythBoundViolation1}. We explained how $r \gtrsim 0.1$, together with constraints on the scalar power spectrum, forces the potential to be too steep to generate sufficient e-folds of inflation for $\Delta \phi \lesssim \mpl/2$ in such a potential.

We hope that this letter removes some of the confusion about the Lyth bound and shows that a measurement of primordial gravity waves with $r \gtrsim 0.1$, if confirmed, truly has the potential to rule out all small-field models of single-field slow-roll inflation.

However, it is still early days yet. Apart from a confirmation of the BICEP2 measurement by other experiments, it is necessary to explore other possible sources which could produce or enhance such a signal, e.g.\ primordial magnetic fields \cite{BICEP:magnetic}, gravity wave production from other sources during inflation \cite{BICEP:axion0,BICEP:axion1,BICEP:axion2,BICEP:axion3,BICEP:axion4,tensorsFromSpectators}, phase transitions in the early universe \cite{BICEP:phaseTrans1,BICEP:phaseTrans2}, non-Bunch-Davies initial conditions \cite{BICEP:nonBD1,BICEP:nonBD2,BICEP:nonBD3,BICEP:nonBD4}, topological defects \cite{BICEP:defects1,BICEP:defects2}, other things that have been overlooked, or any combination of these.

\section*{Acknowledgements}
This work was supported by the Swiss National Science Foundation. We thank Stefano Orani for useful discussions.

\subsection*{Note added on [arXiv:1404.3398v1] and [arXiv:1404.3398v2]}
After the release of this letter, the authors of \cite{LythBoundViolation1,LythBoundViolation2} have written a comment \cite{commentingLetter} expressing their belief that despite our criticism, their results in \cite{LythBoundViolation1,LythBoundViolation2} are correct. In their first version, they provided additional steps in their calculation, which made it more transparent that their original error was in eq.~(7) of \cite{commentingLetter}, where they had set $\int \frac{dy}{y}(\ln y)^{f(y)} = \int \frac{dy}{y} f(y) (\ln y)$.

In the current version (v2) of \cite{commentingLetter}, this step is replaced by a Taylor expansion of the integrand $(k/k_*)^{-1 + \frac{a}{2} + ...}$ around $a=2$, truncated at linear order in $(a-2)$. For values $\lvert a \rvert \lesssim O(10^{-2})$ as required by CMB observations, this approximation is completely inapplicable. As a consequence, their calculation of the integral is still incorrect.

The objections of \cite{commentingLetter} to our criticism are all either based on this integral or unfounded (e.g., we do not use any Taylor expansion of $e^{-2\Delta N}$ to disprove their claim), and our analysis and results remain unchanged.

We do not intend to continue updating this note with future revisions of \cite{LythBoundViolation1,LythBoundViolation2,commentingLetter}. Since they consider single-field slow-roll inflation with $r > 0.1$, a correct calculation will yield $\Delta \phi \gtrsim O(\mpl)$ as one can easily see from our model-independent eq.~\eqref{eq:MAIN}.

\end{document}